\documentclass[pre,twocolumn,showpacs,superscriptaddress,amsmath,amssymb]{revtex4-1}
\usepackage{times}
\usepackage{graphicx}
\usepackage{dcolumn}
\usepackage{subfigure}

\begin{document}

\title{Frequency-difference-dependent stochastic resonance in neural systems}

\author{Daqing Guo}
\email{dqguo@uestc.edu.cn}
\affiliation{The Clinical Hospital of Chengdu Brian Science Institute, MOE Key Lab for Neuroinformation, University of Electronic Science and Technology of China, Chengdu 610054, P. R. China}

\author{Matja\v{z} Perc}
\email{matjaz.perc@uni-mb.si}
\affiliation{Faculty of Natural Sciences and Mathematics, University of Maribor, Koro{\v s}ka cesta 160, SI-2000 Maribor, Slovenia}
\affiliation{CAMTP -- Center for Applied Mathematics and Theoretical Physics, University of Maribor, Mladinska 3, SI-2000 Maribor, Slovenia}

\author{Yangsong Zhang}
\affiliation{The Clinical Hospital of Chengdu Brian Science Institute, MOE Key Lab for Neuroinformation, University of Electronic Science and Technology of China, Chengdu 610054, P. R. China}

\author{Peng Xu}
\affiliation{The Clinical Hospital of Chengdu Brian Science Institute, MOE Key Lab for Neuroinformation, University of Electronic Science and Technology of China, Chengdu 610054, P. R. China}

\author{Dezhong Yao}
\email{dyao@uestc.edu.cn}
\affiliation{The Clinical Hospital of Chengdu Brian Science Institute, MOE Key Lab for Neuroinformation, University of Electronic Science and Technology of China, Chengdu 610054, P. R. China}

\date{\today}

\begin{abstract}
Biological neurons receive multiple noisy oscillatory signals, and their dynamical response to the superposition of these signals is of fundamental importance for information processing in the brain. Here we study the response of neural systems to the weak envelope modulation signal, which is superimposed by two periodic signals with different frequencies. We show that stochastic resonance occurs at the beat frequency in neural systems at the single-neuron as well as the population level. The performance of this frequency-difference-dependent stochastic resonance is influenced by both the beat frequency and the two forcing frequencies. Compared to a single neuron, a population of neurons is more efficient in detecting the information carried by the weak envelope modulation signal at the beat frequency. Furthermore, an appropriate fine-tuning of the excitation-inhibition balance can further optimize the response of a neural ensemble to the superimposed signal. Our results thus introduce and provide insights into the generation and modulation mechanism of the frequency-difference-dependent stochastic resonance in neural systems.
\end{abstract}


\maketitle
\section{Introduction}\label{sec:1}
Cortical neurons operate in noisy environments and display highly irregular firing~\cite{r1}. Recent theoretical studies have revealed the functional importance of noise in modulating neurodynamics. In particular, neurons driven by an appropriate level of stochastic fluctuations have been shown to exhibit several counterintuitive behaviors, such as stochastic resonance (SR)~\cite{r2, r3, r4, r5, r6, r6a1, r7}, inverse SR~\cite{r7a1, r7a2, r7a3, r7a4}, coherence resonance~\cite{r7, r8, r8a1, r9}, synchronization~\cite{r10, r11, r12, r12a1, r12a2}, and energy optimization~\cite{r9a1, r9a2}. Among them, the most studied noise-induced phenomenon is the SR, which originally refers to the enhancement of information transfer in a nonlinear system at an optimal noise level in the presence of a weak periodic signal~\cite{r13, r14}. Remarkably, evidence of SR has been demonstrated in many experimental studies~\cite{r15, r16, r17}, indicating that noise may indeed participate into the signal transduction in neural systems.

The classical SR studies in neuroscience have mainly focused on neural systems driven by an isolated periodic force~\cite{r2, r3, r4, r5, r6, r6a1, r7}. Later investigations have confirmed that the similar SR behaviors can be also observed in neural systems with multiple periodic components~\cite{r18, r19, r20, r21, r22}. For instance, a neuron subject to the mixed periodic signals with harmonic frequencies of a fundamental frequency, shows the maximal response to the fundamental frequency at an intermediate noise level~\cite{r18, r19, r20, r21, r22a1}. This phenomenon is called the ``ghost'' SR (GSR), because it appears at the fundamental frequency missing in the input signals. When input signals are rendered inharmonic by applying a frequency shift equally to all of them, a more generalized GSR behavior can be observed at a linear shift in the response frequency~\cite{r18, r19}. These findings might be biologically critical and yield good agreements with well-designed experiments~\cite{r23, r24, r25}.

Nevertheless, biological neurons may receive more complicated multiple oscillatory signals from various brain regions with different frequencies, ranging from several to hundreds of Hz~\cite{r26, r27, r28}. Mathematically, the temporal superposition of these multiple periodic signals may form a slow envelope modulation signal with the frequency character related to their beat frequencies. It is still not completely established, however, whether the slow-frequency neural information carried by such kind of envelope modulation signal can be stably processed by the brain. Here we show that neural systems can successfully detect the slow-frequency neural information carried by weak envelope modulation signal via the mechanism of SR occurring at the beat frequency. The currently reported SR behavior does not depend on the fundamental frequency, and may thus have important biological applications.

The paper is organized as follows. First, the detailed descriptions of the model are introduced in Sec.~\ref{sec:2}. In Sec.~\ref{sec:3}, we provide the results at the single-neuron level, and then extend them to the population level. Finally, we summarize this work and briefly discuss the biological implications of our findings in Sec.~\ref{sec:4}.

\section{Model}\label{sec:2}
\begin{figure}[!t]
	\includegraphics[width=7.82cm]{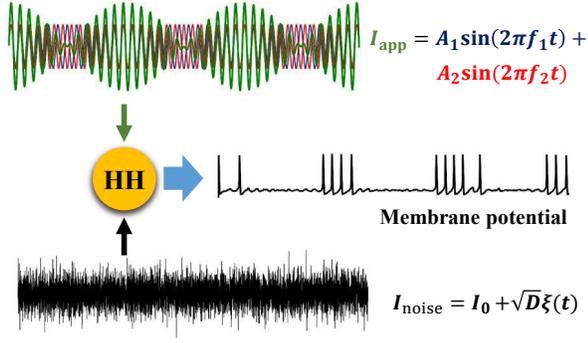}
	\caption{\label{fig:1}(Color online) Schematic presentation of the model. The HH neuron is driven by both the noise current (black) and two sinusoidal signals with frequencies $f_1$ (blue) and $f_2$ (red). The superposition of these two periodic signals forms a relatively slow envelope modulation signal (green) at the beat frequency $f_0$.}
\end{figure}

Let us consider a single neuron driven by two periodic signals with an arbitrary difference in frequency [Fig.~\ref{fig:1}]. The dynamics of the neuron is described by the Hodgkin-Huxley (HH) model, with details as follows~\cite{r29, r30}:
\begin{equation}
\begin{split}
C\frac{dV}{dt}=&-I_{\text{Na}}-I_{\text{K}}-I_{\text{L}}+I_{\text{app}}+I_{\text{noise}}.
\end{split}
\label{eq:1}
\end{equation}
Here $V$ is the membrane potential, $C$ is the membrane capacitance per unit area, and $I_{\text{Na}}=G_{\text{Na}}m_{\infty}^3h(V-E_{\text{Na}})$, $I_{\text{K}}=G_{\text{K}}n^4(V-E_{\text{K}})$ and $I_{\text{L}}=G_{\text{L}}(V-E_{\text{L}})$ represent sodium, potassium and leakage currents through the membrane, respectively. The noise current is modelled as: $I_{\text{noise}}= I_0+\sqrt{D}\xi(t)$, where $I_0$ is the bias current, $\xi(t)$ is the Gaussian white noise with zero mean and unit variance (here the unit of $\xi(t)$ is $\mu$A~ms$^{1/2}$/cm$^2$), and $D$ is a dimensionless parameter denoting the noise intensity. The applied current consists of two periodic signals, which are
\begin{equation}
\begin{split}
I_{\text{app}}=A_1\sin(2\pi f_1 t)+A_2\sin(2\pi f_2 t),
\end{split}
\label{eq:2}
\end{equation}
where $A_1$ and $A_2$ represent signal amplitudes of these two periodic signals, $f_1$ and $f_2$ are their forcing frequencies, and the beat frequency is defined as $f_0=\left|f_2-f_1\right|$. As schematically shown in Fig.~\ref{fig:1}, the superposition of these two periodic signals forms an envelope modulation signal with a slow frequency at $f_0$ (green signal).

In the HH neuron, three gating variables, $x$ ($x=m$, $n$ and $h$), obey the following equation~\cite{r30}:
\begin{equation}
\begin{split}
\frac{dx}{dt}&=\alpha_{x}(1-x)-\beta_{x}x,
\end{split}
\label{eq:3}
\end{equation}
with rate functions given by:
\begin{equation}
\begin{split}
&\alpha_{m}=0.1\frac{25-V}{\exp[(25-V)/10]-1},\\
&\beta_{m}=4\exp{[-V/18]},\\
&\alpha_{n}=0.01\frac{10-V}{\exp[(10-V)/10]-1},\\
&\beta_{n}=0.125\exp[-V/80],\\
&\alpha_{h}=0.07\exp[-V/20],\\
&\beta_{h}=\frac{1}{\exp[(30-V)/10]+1}.
\end{split}
\label{eq:4}
\end{equation}
In simulations, we use the following parameters for the HH neuron~\cite{r30}: $C=1$ $\mu$F/cm$^2$, $G_{\text{Na}}=120$ ms/cm$^2$, $E_{\text{Na}}=115$ mV, $G_{\text{K}}=36$ ms/cm$^2$, $E_{\text{K}}=-12$ mV, $G_{\text{L}}=0.3$ ms/cm$^2$, $E_{\text{L}}=10$ mV. Unless otherwise noted, we set $I_0=1$~$\mu$A/cm$^2$, and $A_1=A_2=0.6$~$\mu$A/cm$^2$. In the absence of noise, the applied current is too weak to excite the HH neuron for different frequency combinations considered in this work.  The model is integrated using the Euler-Maruyama method with a time step $h=0.01$~ms~\cite{r31}. All computer codes will be available to download from ModelDB (\url{https://senselab.med.yale.edu/ModelDB/}).

\section{Results}\label{sec:3}
\begin{figure}[!t]
	\includegraphics[width=8.5cm]{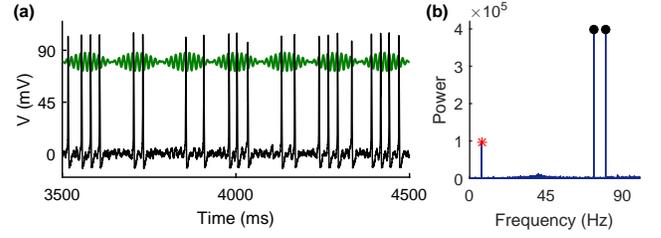}
	\caption{\label{fig:2}(Color online) A typical example of the stochastic oscillation in the HH neuron, with parameters: $f_1=73$~Hz, $f_2=80$~Hz and $D=2.5$. (a) The trace of the membrane potential $V$ (black)  and the envelope modulation signal (green). For better visualizing, the amplitude of envelope modulation signal is magnified 5 times with an offset of 80~$\mu$A/cm$^2$. (b) PSD of the spike train (50 seconds). In (b), red asterisk represents the power at the beat frequency $f_0=7$~Hz, whereas black circles denote the powers at two forcing frequencies $f_1$ and $f_2$.}
\end{figure}

We first set out to examine whether the oscillation characteristic at the beat frequency can be exhibited in the spike train generated by the HH neuron. Figure~\ref{fig:2} shows an example of the stochastic oscillation of the HH neuron, with forcing frequencies $f_1=73$~Hz and $f_2=80$~Hz. Due to the existence of stochastic fluctuations, the HH neuron displays irregular firing, but its firing pattern roughly matches with the waveform of the superimposed signal [Fig.~\ref{fig:2}(a)]. By further estimating the power spectral density (PSD) of spike train using the fast Fourier transform, we identify three main power peaks located at the beat frequency $f_0=7$~ Hz as well as two forcing frequencies $f_1=73$~Hz and $f_1=80$~Hz [Fig.~\ref{fig:2}(b)], respectively. Besides, several other power peaks located at multiples of these two forcing frequencies can be also observed (data not shown). These findings indicate that neural information at the beat frequency carried by the weak envelope modulation signal can be successfully detected in the spike train of the HH neuron.

\begin{figure*}[!t]
	\includegraphics[width=16.5cm]{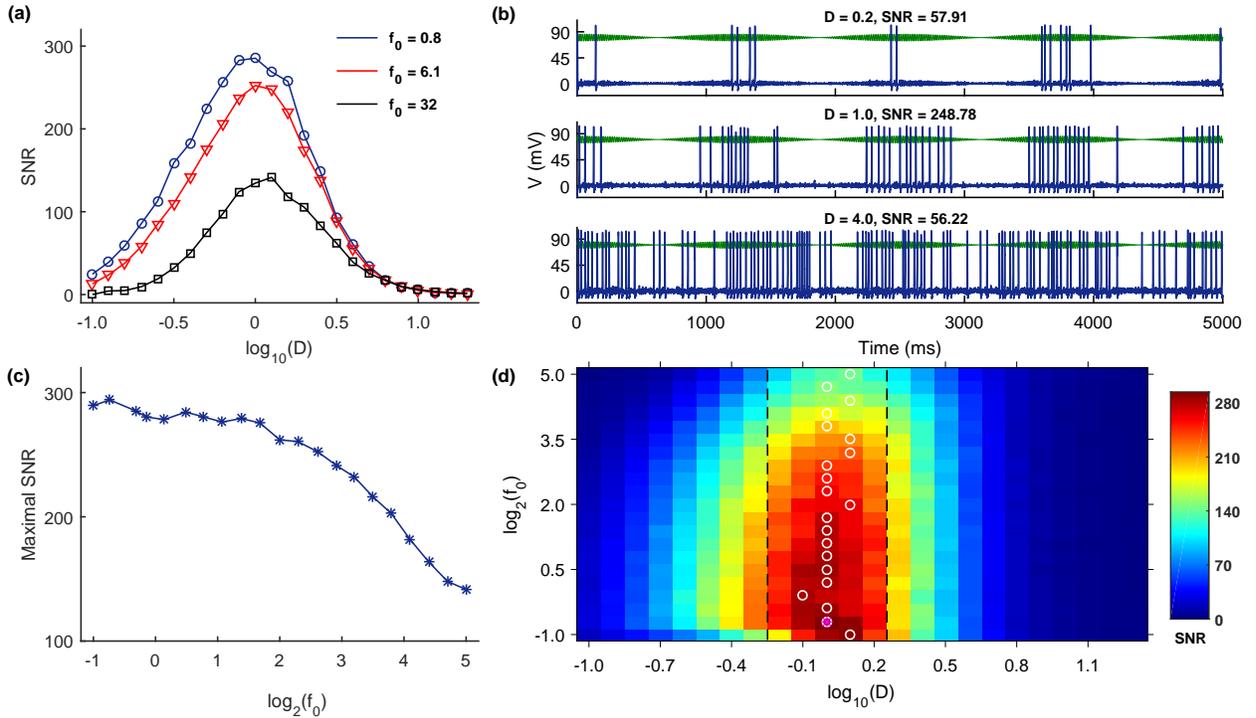}
	\caption{\label{fig:3}(Color online) Response of a single HH neuron to different values of $D$ and $f_0$. (a) SNR versus $D$ for different beat frequencies. (b) Typical trace of the membrane potential $V$ (blue) and the corresponding envelope modulation signal (green) at different noise levels. For better visualizing, the amplitude of envelope modulation signal is magnified 5 times with an offset of 80~$\mu$A/cm$^2$.	(c) Maximal SNR as a function of $f_0$. (d) SNR in the $(D,f_0)$ panel. In (d), each white circle refers to the maximal SNR point for corresponding $f_0$, the magenta asterisk denotes the maximal SNR point in the $(D,f_0)$ panel, and the region between two black dash lines denotes the optimal noise regime. The unit of the beat frequency $f_0$ is Hz. In all simulations, we set $f_1=73$~Hz and $f_2=f_1+f_0$. In (b), the beat frequency is fixed at $f_0=0.8$~Hz, and three noise intensities are: $D=0.2$, $D=1.0$ and $D=4.0$, respectively.}
\end{figure*}

We next ask whether the SR-type behavior can be observed at the beat frequency in a single neuron. To address this, we quantify the capability of information transfer by using the signal-to-noise ratio (SNR). In the present study, the SNR is estimated from the PSD, defined as~\cite{r3, r7}: $\text{SNR}=[S(f_0)-N(f_0)]/N(f_0)$, where $S(f_0)$ is the power at the beat frequency $f_0$ and $N(f_0)$ is the averaged power at nearby frequencies. For each experimental setting, we carry out 50 realizations of simulations and report the averaged SNR as the final result.

Figure~\ref{fig:3}(a) shows the SNR value versus the noise intensity $D$ for different beat frequencies. With increasing $D$, each SNR curve first rises and then drops, and the maximal SNR value is achieved at an intermediate noise level. Consistently, we observe that the membrane potential of the neuron matches well with the waveform of superimposed signal at an intermediate noise level, and exhibits a poor performance when the neuron driven by either low or high level of stochastic fluctuations [see Fig.~\ref{fig:3}(b)]. These findings demonstrate the occurrence of SR and indicate that beat-frequency related neural information carried by the weak envelope modulation signal can be well detected with the help of noise. Since this type of SR occurs at the beat frequency, we term it as the frequency-difference-dependent SR in this study. Interestingly, we find that each SNR curve shows the maximal response to its corresponding beat frequency $f_0$ at almost the similar noise level [Fig.~\ref{fig:3}(a)]. For a fixed forcing frequency $f_1$, this observation suggests that the optimal noise-enhanced region might be not impacted by the beat frequency in the frequency-difference-dependent SR.

\begin{figure}[!b]
	\includegraphics[width=8.5cm]{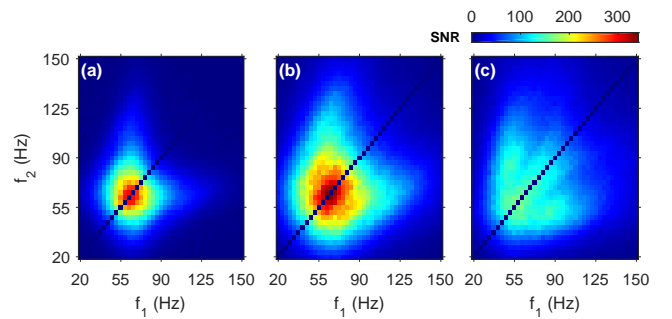}
	\caption{\label{fig:4}(Color online) Dependence of the SNR on two forcing frequencies $f_1$ and $f_2$ at different levels of neuronal noise. From (a)-(c), three noise intensities considered here are: $D=0.4$, $D=1.0$ and $D=2.5$, respectively. }
\end{figure}

To explore the effects of beat frequency on the performance of frequency-difference-dependent SR, we calculate the maximal SNR at the corresponding optimal noise intensity for different $f_0$. As we see in Fig.~\ref{fig:3}(c), the maximal SNR gradually decreases with the increase in $f_0$, suggesting that the HH neuron may show a better performance at a relatively smaller beat frequency. By further presenting the SNR value in the $(D,f_0)$ panel, we observe that strong neural response mainly appears at the small beat frequency range within optimal noise-enhanced regime [Fig.~\ref{fig:3}(d)]. These results provide consistent evidence that the performance of frequency-difference-dependent SR is especially sensitive to small beat frequency. In the brain, we presume that neurons may use this mechanism to discriminate multiple oscillatory signals with fine frequency distinctions.

In reality, the performance of frequency-difference-dependent SR is also significantly influenced by the absolute sizes of two forcing frequencies. In Figs.~\ref{fig:4}(a)-\ref{fig:4}(c), we plot the SNR value in the $(f_1,f_2)$ panel at three different levels of neuronal noise. For each noise intensity, the neuron responds optimally to the superimposed signal at a special frequency range (40-90 Hz) within the gamma band. This phenomenon is the so-called frequency sensitivity, which has been reported in neural systems and might be due to the cooperation of the intrinsic oscillation of neurons and the external periodic input signals~\cite{r5, r6, r7}. Since the input current has as sinusoidal form, both the mean and the variance do not change as the frequency of the current changes, and accordingly, the same effects would be observed if normalizing the injected inputs by the input frequency. Furthermore, we observe that such frequency sensitivity is shaped by the neuronal noise [see Figs.~\ref{fig:4}(a)-\ref{fig:4}(c)]. At an intermediate level of noise, the HH neuron exhibits a wider frequency-sensitivity range than those for both low and high noise levels. We highlight these findings because neural oscillations in the gamma band have widely observed in the brain, and are believed to play important role in enhancing information transmission between groups of neurons~\cite{r31a1, r31a2} and to be associated with many higher cognitive tasks~\cite{r32, r33, r34, r35}.

\begin{figure}[!t]
	\includegraphics[width=8cm]{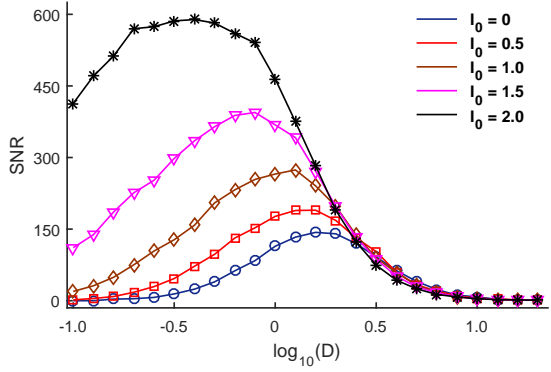}
	\caption{\label{fig:5}(Color online) Response of a single HH neuron to different values of $D$ and $I_{0}$, with $f_1=73$~Hz and $f_2=3$~Hz. Five bias currents considered here are:  $I_0=0$, 0.5, 1.0, 1.5, and 2.0~$\mu$A/cm$^2$, from bottom to top, respectively.}
\end{figure}

To examine the effects of the bias current on the performance of frequency-difference-dependent SR, we plot the SNR value as a function of $D$ for different values of $I_{0}$. It should be noted that, for all bias currents considered here, the applied current is maintained to be subthreshold. As shown in Fig.~\ref{fig:5}, each SNR curve displays a bell-shaped curve, further implying that the frequency-difference-dependent SR is an inherent property of the HH neuron driven by the subthreshold stimulus. Theoretically, with the increasing of the bias current $I_{0}$, the membrane potential of the HH neuron is pushed close to its firing threshold, thus requiring a relatively low level of neuronal noise to trigger action potential. As a result, we observe that the SNR curve is shift to the top left as $I_0$ grows, and a stronger maximal response to the superimposed signal is achieved at a relatively weaker optimal noise level for the HH neuron.

\begin{figure}
	\includegraphics[width=8.5cm]{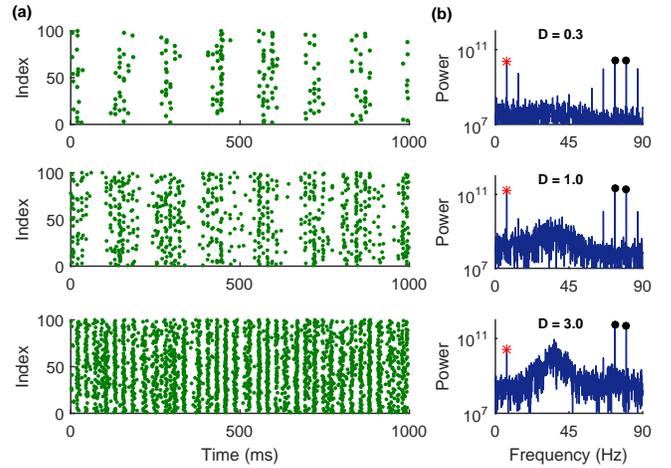}
	\caption{\label{fig:6}(Color online) Typical response of the network at different noise levels. (a) Spike raster and (b) the corresponding PSD of the network activities. From top to bottom, three noise intensities are: $D=0.3$, $D=1.0$ and $D=3.0$. Other parameters are set as $f_1=73$~Hz, $f_2=80$~Hz, $w_{\text{E}}=0.03$~ms/cm$^2$, and $w_{\text{I}}=0.15$~ms/cm$^2$ in simulations. In (b), red asterisk represents the power at the beat frequency $f_0=7$~Hz, whereas black circles denote the powers at two forcing frequencies $f_1$ and $f_2$.}
\end{figure}

So far, we have identified the occurrence of frequency-difference-dependent SR at the single-neuron level. A natural question is whether the similar SR can be also observed at the population level. To answer this, we establish a random neuronal network composed of 80 excitatory and 20 inhibitory neurons with a connection density $p=0.1$. The dynamic of neurons is simulated using the HH model. For each neuron, we also incorporate the conductance-based synaptic current into the model, given as: $I_{\text{syn}}= g_{\text{E}} (V_{\text{E}}-V) + g_{\text{I}} (V_{\text{I}}-V)$, where $V_{\text{E}}=60$~mV and $V_{\text{I}}=-20$~mV are excitatory and inhibitory synaptic reversal potentials, and $g_{\text{E}}$ and $g_{\text{I}}$ are their corresponding synaptic conductances. Whenever a neuron receives a presynaptic spike, its synaptic conductance is updated according to, $g_{\text{E}} \leftarrow g_{\text{E}}+w_{\text{E}}$ for an excitatory spike and $g_{\text{I}} \leftarrow g_{\text{I}}+w_{\text{I}}$ for an inhibitory spike. In other time, these two synaptic conductances decay in an exponential manner with fixed time constants $\tau_{\text{E}}=5$~ms and $\tau_{\text{I}}=10$~ms. Parameters $w_{\text{E}}$ and $w_{\text{I}}$ represent the synaptic strengths of excitatory and inhibitory synapses. The mean firing rate $s(t)$ measured in 0.1~ms bin size is employed to estimate the PSD of network activities, which is further utilized to calculate the SNR at the beat frequency.

Figure~\ref{fig:6}(a) illustrates three typical spike raster diagrams for different noise intensities. When the noise level is low, neurons in the network generate few scattered spikes due to weak stochastic fluctuations. In this case, a part of neural information is lost during the transmission, leading to a small power peak at the beat frequency [Fig.~\ref{fig:6}(b), top panel]. As the noise intensity increases, stochastic fluctuations from noise start to affect neuronal firing. For an appropriate noise level, the collective firing of neurons responds well to the waveform of superimposed signal. As a consequence, a large power peak can be observed at the beat frequency [Fig.~\ref{fig:6}(b), middle panel]. However, if the neuronal noise is too strong, the applied current is almost drowned in noise, and neuronal firing is determined by both strong noise current and synaptic interaction. Under this condition, the network exhibits fast gamma oscillations and neural information carried by the low-frequency envelope modulation signal cannot be directly read from the collective firing of neurons [Fig.~\ref{fig:6}(b), bottom panel]. Consist with previous findings, the association of increased information retrieved form the network with increased gamma power might support the notion of gamma oscillations playing a role in information processing for networks with strong synaptic interactions~\cite{r38a1}.

\begin{figure}
	\includegraphics[width=8.5cm]{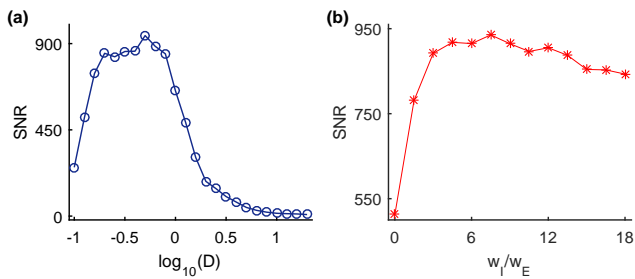}
	\caption{\label{fig:7}(Color online) Population response under different conditions. (a) The SNR value as a function of $D$, with fixed synaptic strengths $w_{\text{E}}=0.03$~ms/cm$^2$ and $w_{\text{I}}=0.15$~ms/cm$^2$. (b) Dependence of SNR on the relative strength of excitatory and inhibitory synapses $w_{\text{E}}/w_{\text{I}}$, with parameters $w_{\text{E}}=0.03$~ms/cm$^2$ and $D=0.55$. In simulations, we set two forcing frequencies as $f_1=73$~Hz, $f_2=80$~Hz.}
\end{figure}

To quantitatively validate the above observation, we further illustrate the relationship between the SNR value and the noise intensity $D$ in Fig.~\ref{fig:7}(a). As expected, a bell-shaped SNR curve is observed with the increase of $D$, indicating that the network shows the best response to the superimposed signal at the beat frequency for an optimal noise level. However, due to stochastic fluctuations introduced by synaptic interaction, we find that the optimal noise-enhanced region at the population level is shift toward lower noise intensity [Fig.~\ref{fig:3} and Fig.~\ref{fig:7}(a)]. Such finding demonstrates that the frequency-difference-dependent SR can indeed appear at the population level. More importantly, the maximal SNR value at the population level is much larger than that at the single-neuron level [Fig.~\ref{fig:3} and Fig.~\ref{fig:7}(a)], suggesting that the collective firing of neurons might be more efficient to detect and transmit the low-frequency neural information carried by the weak envelope modulation signal. Note that this finding might be especially suitable for a single neuron in the network which does not generally fire on every periodic cycle due to heterogeneous feedback inhibition~\cite{r38a2}, even when the two driven frequencies fall into the gamma band.

Finally, we also find that the relative strength of excitatory and inhibitory synapses plays a critical role in regulating the performance of frequency-difference-dependent SR at the  population level [Fig.~\ref{fig:7}(b)]. For a fixed noise level, our results reveal that the optimal network response to the envelope modulation signal at the beat frequency is achieved at an intermediate relative strength. From the theoretical perspective, this is because a fine balance between excitation and inhibition prevents excessive neuronal firing and contributes to network stability, thus supporting stable and robust weak signal detection and transmission.

\section{Discussion}\label{sec:4}
In summary, we have examined the stochastic dynamics of neural systems driven by two periodic signals with arbitrary difference in frequency. We observed that the frequency-difference-dependent SR occurs at both the single-neuron and population levels. Our simulations showed that the performance of frequency-difference-dependent SR does not only relies on the relative size of beat frequency, but is also impacted by the absolute sizes of two forcing frequencies. By analyzing the frequency-sensitivity of neurons, we identified a special frequency range (40-90 Hz) within the gamma band. At an intermediate noise level, the neuron shows relatively strong response to external periodic signals when their frequencies fall into this special range. This finding is of importance because gamma neural oscillations have been believed to modulate and enhance signal transmission~\cite{r31a1, r31a2}, and have been linked to many higher cognitive tasks~\cite{r32, r33, r34, r35}. Remarkably, we found that population response of neural ensembles is more efficient than that of a single neuron to detect the neural information carried by the envelope modulation signal at the beat frequency. Further investigations reveled that a fine excitation-inhibition balance can improve the network response to the envelope modulation signal at the beat frequency. These results shed insights into the functional roles of stochastic noise in promoting the signal transduction for the beat-frequency related neural information.

Dynamical response of neurons to noisy oscillatory inputs is fundamental for neural information processing~\cite{r36, r37, r38}. Our results confirm that neural systems can also respond to the weak frequency-difference information through the mechanism of SR. This finding might offer important biological implications, because realistic neurons are often simultaneously driven by multiple oscillatory signals with different frequencies~\cite{r26, r27, r28}. After a long time of evolution, it is reasonable to suppose that our brain might have the abilities to use neuronal noise to achieve stable transmission for frequency-difference information, which can be further used by the brain to perform higher cognitive tasks. We hope that predictions from our model investigation can inspire testable hypotheses for electrophysiological experiments in the future. Further work on this topic includes investigating the neuronal response to multiple suparthreshold periodic signals, and investigating possibles roles of frequency-difference-dependent SR in regulating complicated neurodynamics~\cite{r11, r39, r40}.

\begin{acknowledgments}
We sincerely thank Dr. Mingming Chen for valuable comments on an early version of this manuscript. This research was supported by the National Natural Science Foundation of China (Grants Nos. 81571770, 61527815, 81401484 and 81330032) and the Slovenian Research Agency (Grants P5-0027 and J1-7009).
\end{acknowledgments}

\end{document}